\documentclass[twolecolumn]{elsart3p}

\usepackage{amssymb}
\usepackage[dvips]{graphicx}
\usepackage{bm}
\usepackage{amsmath}

\journal{Physica C}

\begin{document}

\begin{frontmatter}



\title{Effect of uniaxially anisotropic Fermi surface on the
quasiparticle scattering inside a vortex core in unconventional
superconductors}


\author[AA,CC,DD]{Yoichi Higashi\corauthref{cor1}},\,
\ead{higashiyoichi@ms.osakafu-u.ac.jp}\,
\author[BB,DD,EE]{Yuki Nagai},\,
\author[BB,DD,EE]{Masahiko Machida},\,
\author[CC,DD]{Nobuhiko Hayashi}
\address[AA]{Department of Mathematical Sciences, Osaka Prefecture University, 1-1 Gakuen-cho, Naka-ku, Sakai 599-8531, Japan}

\address[BB]{CCSE, Japan Atomic Energy Agency, 5-1-5 Kashiwanoha, Kashiwa, Chiba 277-8587, Japan}

\address[CC]{NanoSquare Research Center (N2RC), Osaka Prefecture University, 1-2 Gakuen-cho, Naka-ku, Sakai 599-8570, Japan}

\address[DD]{CREST(JST), 4-1-8 Honcho, Kawaguchi, Saitama 332-0012, Japan}

\address[EE]{TRIP(JST), 5 Sanban-cho, Chiyoda-ku, Tokyo 110-0075, Japan}

\corauth[cor1]{Corresponding author.
N2RC, Osaka Prefecture University, C10 Bldg., 1-2 Gakuen-cho, Naka-ku, Sakai 599-8570, Japan
Tel.: +81-72-254-9829 ; fax: +81-72-254-8203}
\begin{abstract}
We theoretically study the dependence of the quasiparticle (QP) scattering rate $\varGamma$ on the uniaxial anisotropy of a Fermi surface with changing the magnetic field angle $\alpha_{\rm M}$. 
We consider the QP scattering due to the non-magnetic impurities inside a single vortex core.
The field-angle dependence of the quasiparticle scattering rate $\varGamma(\alpha_{\rm M})$ is sensitive to the sign-change of the pair potential.
We show that with increasing the two dimensionality of the system, $\varGamma(\alpha_{\rm M})$ reflects more clearly whether there is the sign-change in the pair potential.
\end{abstract}
\begin{keyword}
Unconventional superconductors \sep
Field-angle dependent measurement \sep
Quasiparticle scattering rate \sep
Phase sensitive probe
\PACS 74.20.Rp \sep 74.25.Op \sep 74.25.nn
\end{keyword}
\end{frontmatter}

\section{Introduction}

In anisotropic superconductors,
the pair potential has the anisotropy in the $\bm{k}$-space.
For example, cuprate superconductors and the heavy fermion superconductor CeCoIn$_5$ are experimentaly identified as $d_{x^2-y^2}$-wave superconductors \cite{tsuei, field-angle}.
These superconductors have the zero of the pair potential (i.e. gap node) in $\pi/4$ direction measured from the $a$-axis.
When the azimuthal angle goes across the gap node direction, the pair potential changes its sign.
It is of great importance to elucidate the pair potential symmetry because it reflects the characteristics of the pairing interaction.
Therefore, many experiments have been carried out in order to investigate the pairing symmetry.
The field-angle resolved thermal conductivity and specific heat measurments are powerful techniques, which can detect the anisotropy of the pair potential amplitude \cite{field-angle}.
However, they are not phase-sensitive probe.
Information on the sign change of the pair potential is crucial for clarifying the unconventional Cooper pairing (e.g., $d_{x^2-y^2}$-wave pair).
There are a few phase-sensitive probes \cite{tsuei, hanaguri}, but an easier way to detect the sign change of the pair potential is required.

In a previous paper,
we reported that the field-angle ($\alpha_{\rm M}$) dependence of the flux-flow resistivity is phase-sensitive in the case of an isotropic spherical Fermi surface (FS) \cite{higashi}.
In this paper, the dependence of the quasiparticle (QP) scattering rate $\varGamma$ on the uniaxial anisotropy of a FS is theoretically investigated under rotating the applied magnetic field $\bm{H}$.
We take into account the electronic structure reflecting the crystal structure anisotropy by means of changing the dimensionality of the system.
Then we consider a spheroidal FS, which is closer to realistic systems than a spherical FS (see Fig.~\ref{Fig.1}).
As a result of our calculation,
we show that with increasing the uniaxial anisotropy of the system, $\varGamma$ depends more strongly on $\alpha_{\rm M}$
and $\varGamma(\alpha_{\rm M})$ reflects more clearly the Cooper pairing symmetry including the sign-change of the pair potential.
\section{Formulation}
We consider a single vortex core in which non-magnetic impurities are distributed randomly.
$\varGamma$ is obtained by calculating the imaginary part of the impurity-induced self energy $\varSigma$ on the basis of the quasiclassical theory of superconductivity \cite{serene}.
$\varSigma$ appears as an energy shift of the vortex bound state in the denominator of the regular quasiclassical Green's function.
The quasiclassical Green's functions are obtained analytically in the vicinity of a vortex core by using the Kramer-Pesch approximation \cite{KP}.
We treat the impurity scattering by using the $t$-matrix approximation \cite{thuneberg}.
We assume an $s$-wave scattering potential and consider the Born limit.
The pair potential is given as $\Delta({\bm r},{\bm k}_{\rm F})=f({\bm r}) \Delta_0 d({\bm k}_{\rm F}) e^{i\phi}$,
where $f({\bm r})$ denotes the spatial variation of the pair potential,
$\Delta_0$ is its bulk amplitude and $\phi$ is the azimuthal angle.
$f(\vert {\bm r} \vert \rightarrow \infty)=1$ and $f({\bm r}=0)=0$.
The center of the vortex line is located at ${\bm r}=0$.
$d({\bm k}_{\rm F})$ represents the anisotropy of the pair potential in the ${\bm k}$-space.
${\bm k}_{\rm F}$ is the Fermi wave vector (i.e. the position on a spheroidal FS).
The scattering rate $\varGamma(\varepsilon)$ for the vortex bound states with energy $\varepsilon$ is given by \cite{nagai-kato}
\begin{eqnarray}
\frac{\varGamma(\varepsilon)}{\varGamma_{\rm n}}
&=&
\frac{\pi}{2}  
\Bigg\langle
\Bigg\langle
     \bigl(1-\mathop{\mathrm{sgn}}\nolimits[d({\bm k}_{\rm F})d({\bm k}^\prime_{\rm F})] \cos\Theta  \bigr) 
\nonumber \\
& & { } \qquad   \times
\frac{1}{\vert\sin\Theta\vert}\frac{\vert {\bm v}_{\rm F \perp}({\bm k}^\prime_{\rm F}) \vert}{\vert \bm{v}_{\rm F \perp}({\bm k}_{\rm F}) \vert}\frac{\vert d({\bm k}_{\rm F}) \vert}{\vert d({\bm k}^\prime_{\rm F})\vert}
\nonumber \\
& & { } \qquad
e^{-u(s_0,{\bf k}_{\rm F})}e^{-u(s^\prime_0,{\bf k}^\prime_{\rm F})}
\Bigg\rangle_{\rm FS^\prime}
\Bigg\rangle_{\rm FS},
\nonumber \\
\label{eq:1}
\end{eqnarray}
where $\varGamma_{\rm n}$ is the scattering rate in the normal state.
The brackets $\langle \cdots \rangle_{\mathop{\mathrm{FS}} }$ and
$\langle \cdots \rangle_{\mathop{\mathrm{FS}}^\prime}$ mean the averages on a FS
with respect to ${\bm k}_{\rm F}$ and ${\bf k}^\prime_{\rm F}$ respectively, like 
$\langle \cdots \rangle_{\mathop{\mathrm{FS}}}\equiv(1/\nu_0)\int dS_{\rm F}({\bm k}_{\rm F})/\vert {\bm v}_{\rm F}({\bm k}_{\rm F}) \vert \cdots$
with $dS_{\rm F}$ being an area element on a FS
and $\nu_0=\int dS_{\rm F}({\bm k}_{\rm F})/\vert {\bm v}_{\rm F}({\bm k}_{\rm F}) \vert$.
The QP is scattered from the initial position ${\bm k}_{\rm F}$ to the final one ${\bm k}^\prime_{\rm F}$ on the spheroidal FS.
$\vert {\bm v}_{\rm F \perp} ({\bm k}_{\rm F})\vert$ and $\vert {\bm v}^\prime_{\rm F \perp}({\bm k}^\prime_{\rm F}) \vert$ reflect the uniaxial anisotropy of the FS.
The index g$\perp$hmeans the component projected onto the plane normal to ${\bm H}$.
The coherence length in the bulk is defined by $\xi_0=v_{\rm F \perp}/\pi \Delta_0$ with $v_{\rm F \perp} \equiv \langle \vert {\bm v}_{\rm F \perp}({\bm k}_{\rm F}) \vert \rangle_{\mathop{\mathrm{FS}}}$.
The details of Eq.~(\ref{eq:1}) are discussed in Ref. \cite{higashi, nagai-kato}.

In this study, we consider the case where ${\bm H}$ is rotated in the $a - b$ plane perpendicular to the $k_z$-axis.
The relations between the coordinates fixed to $\bm{H}$ (vortex coordinates) and those fixed to the crystal axes (crystal coordinates) are described in detail in the case of spherical FS in Ref. \cite{higashi}.
In a two dimensional system,
when $\bm{H}$ is applied within the $a - b$ plane,
a distorted vortex core is formed.
However, we neglect the vortex core distortion here and assume an isotropic vortex core.
We set the variation of the pair potential in the plane normal to ${\bm H}$, like $f({\bm r})=\tanh(|{\bm r}|/\xi_0)$.

\begin{figure}[t]
\begin{center}
\includegraphics[width = 40mm]{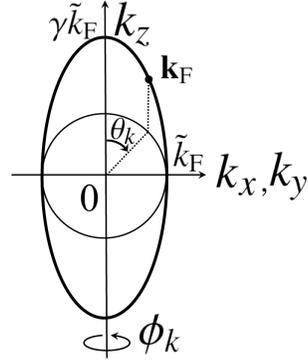}
\caption{
Cross section view of a spheroidal FS along a certain $\phi_k$ with the anisotropy $\gamma$.
The vertical axis is $k_z$-axis and the horizontal axis is $k_x$ or $k_y$-axis.
The position on the spheroidal FS is characterized by the parameter $(\phi_k, \theta_k)$ on a spherical shell
with the radius $\tilde{k}_{\rm F}$.
}
\end{center}
\label{Fig1}
\end{figure}

We consider a spheroidal FS uniaxially extended to the $k_z$-direction (see Fig.~\ref{Fig1}).
For a spheroidal FS,
the area element is given by
\begin{equation}
dS_{\rm F}=\tilde{k}^2_{\rm F} \sqrt{\cos^2 \theta_k+\gamma^2 \sin^2 \theta_k}\sin \theta_k d\phi_k d \theta_k,
\end{equation}
where the uniaxial anisotropy parameter is defined as $\gamma \equiv \sqrt{m_c/m_{ab}}$.
The mass $m_{ab}$ and $m_c$ characterize the charge transport within the $a - b$ plane and in the $c$-axis, respectively.
$\tilde{k}_{\rm F}$ is the radius of a spherical shell (see Fig.~\ref{Fig1}).
We assume the system is isotropic in the $a - b$ plane.
$\phi_k$ ($\theta_k$) is the azimuthal (polar) angle on the spherical shell.
The position on the spheroidal FS $\bm{k}_{\rm F}$ is uniquely identified by the position on this spherical shell ($\phi_k$, $\theta_k$) (Fig.~\ref{Fig1}).

On the spheroidal FS,
the Fermi velocity is expressed by \cite{graser}
\begin{equation}
{\bm v}_{\rm F}({\bf k}_{\rm F})=\tilde{v}_{\rm F}(\hat{{\bm a}} \cos \phi_k \sin \theta_k + \hat{{\bm b}} \sin \phi_k \sin \theta_k +\hat{{\bm c}}(1/\gamma) \cos \theta_k),
\end{equation}
where $\hat{{\bm a}}, \hat{{\bm b}}, \hat{{\bm c}}$ are orthogonal unit vectors characterizing the crtystal axes.
$\tilde{v}_{\rm F}$ is the Fermi velocity of in the $a - b$ plane.
After transforming the coordinate system from the crystal coordinates into the vortex coordinates,
we can get the relations 
\begin{equation}
|{\bm v}_{{\rm F}\bot} ({\bm k}_{\rm F})|
=
\tilde{v}_{\rm F}
\sqrt{
\frac{1}{\gamma^2}\cos^2\theta_k + \sin^2(\phi_k-\alpha_{\rm M}) \sin^2\theta_k
},
\end{equation}
\begin{equation}
\cos\theta_v ({\bm k}_{\rm F})
=
-\frac{1}{\gamma}\frac{ \tilde{v}_{\rm F} }{ |{\bm v}_{{\rm F}\bot} ({\bm k}_{\rm F})| }
\cos\theta_k,
\end{equation}
\begin{equation}
\sin\theta_v({\bm k}_{\rm F})
=
\frac{ \tilde{v}_{\rm F} }{ |{\bm v}_{{\rm F}\bot} ({\bm k}_{\rm F})| }
\sin(\phi_k-\alpha_{\rm M}) \sin\theta_k,
\label{eq:2}
\end{equation}
where
${\bm k}_{\rm F}
=
\tilde{k}_{\rm F} (
\hat{{\bm a}} \cos\phi_k \sin\theta_k
+
\hat{{\bm b}} \sin\phi_k \sin\theta_k
+
\hat{{\bm c}} \gamma \cos\theta_k
)$.
$\theta_v({\bm k}_{\rm F})$ is the angle of the QP trajectory measured from the ${\bm a}_{\rm M}$-axis.
The ${\bm a}_{\rm M}$-axis is unit vector spanning the plane normal to ${\bm H}$.
Here, $\bm{H}=|\bm{H}|(\hat{\bm{a}} \cos \alpha_{\rm M} + \hat{\bm{b}} \sin \alpha_{\rm M})$.
All quantities are represented in the coordinate system fixed to the crystal axes.
Therefore we can calculate the $\alpha_{\rm M}$ dependence of $\varGamma$ in the same way as in the case of a spherical FS \cite{higashi}.
\section{Results}
\begin{figure}[t]
 \begin{center}
   \begin{tabular}{p{80mm}p{80mm}}
     \resizebox{80mm}{!}{\includegraphics{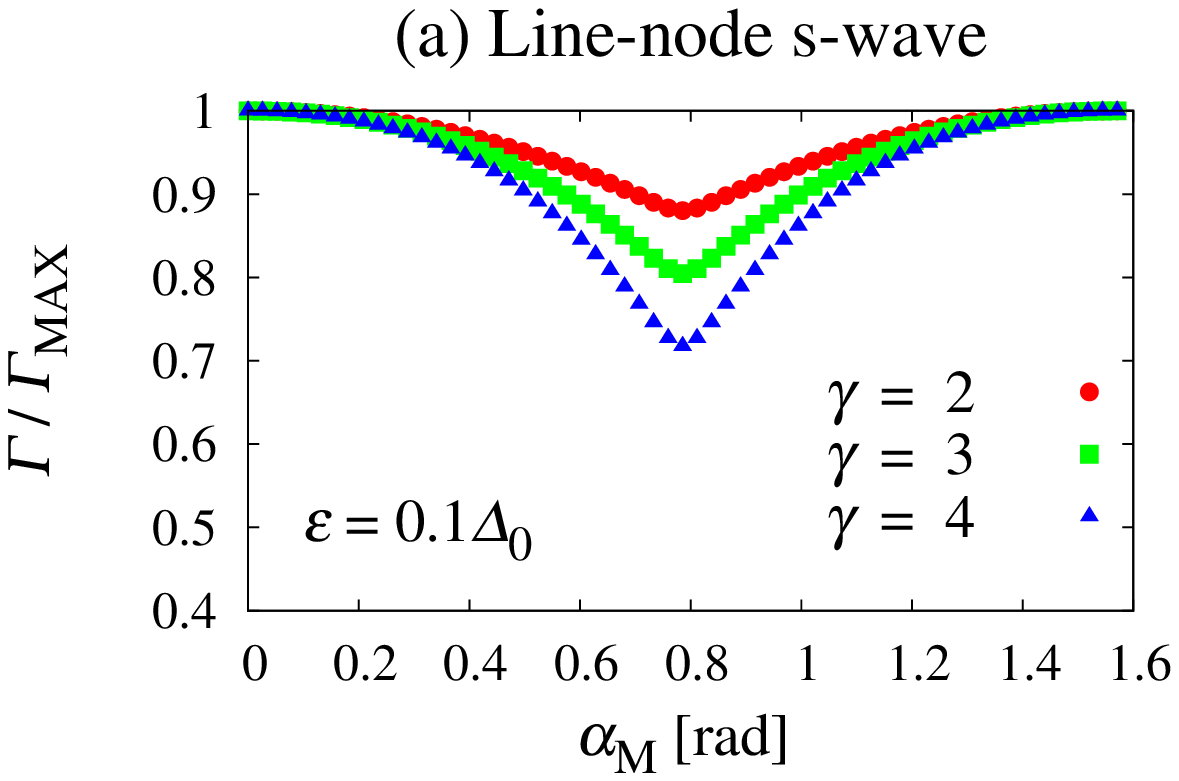}}\\
     \resizebox{80mm}{!}{\includegraphics{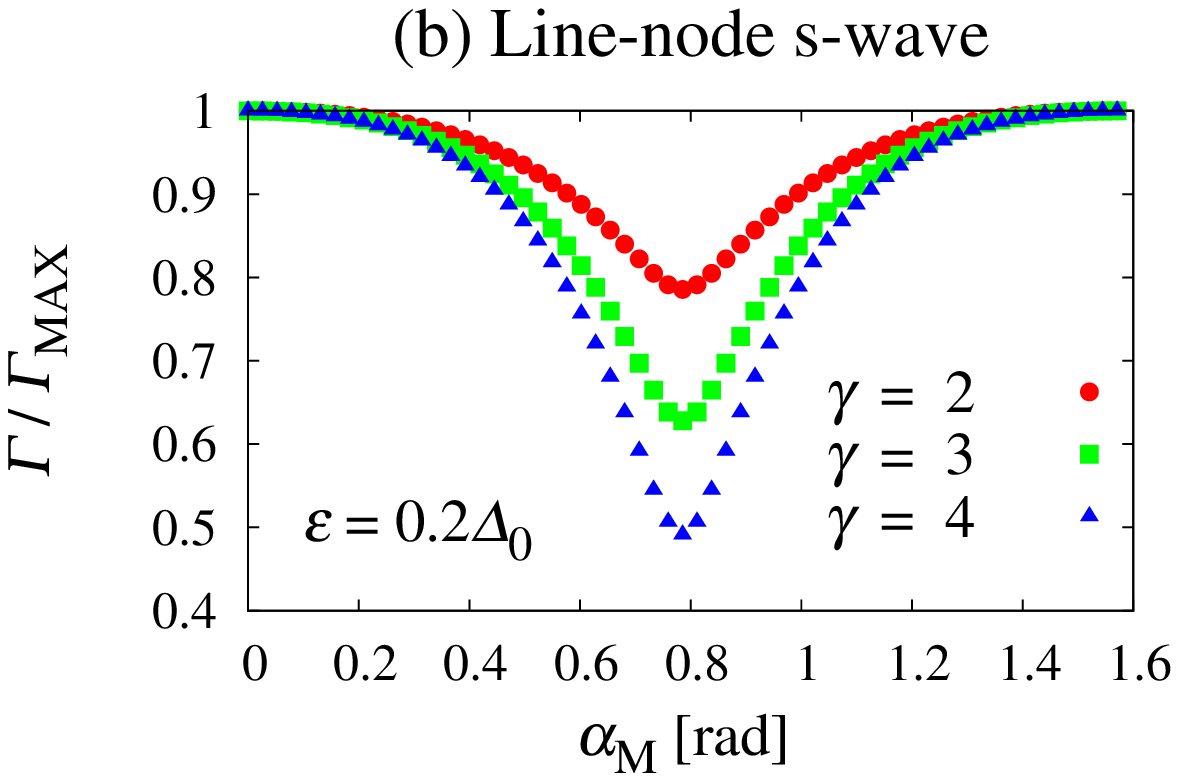}}
   \end{tabular}
 \end{center}
 \caption{Plot of the QP scattering rate $\varGamma$ vs.\ the applied field angle $\alpha_{\rm M}$
in the case of the line-node $s$-wave pair for (a) the QP energy $\varepsilon=0.1\Delta_0$ and (b) $\varepsilon=0.2\Delta_0$.
Each curve is plotted for different anisotropy $\gamma$.
The vertical axis is normalized by maximum value for each curve.}
 \label{Fig.2}
\end{figure}
\begin{figure}[t]
 \begin{center}
    \begin{tabular}{p{80mm}p{80mm}}
      \resizebox{80mm}{!}{\includegraphics{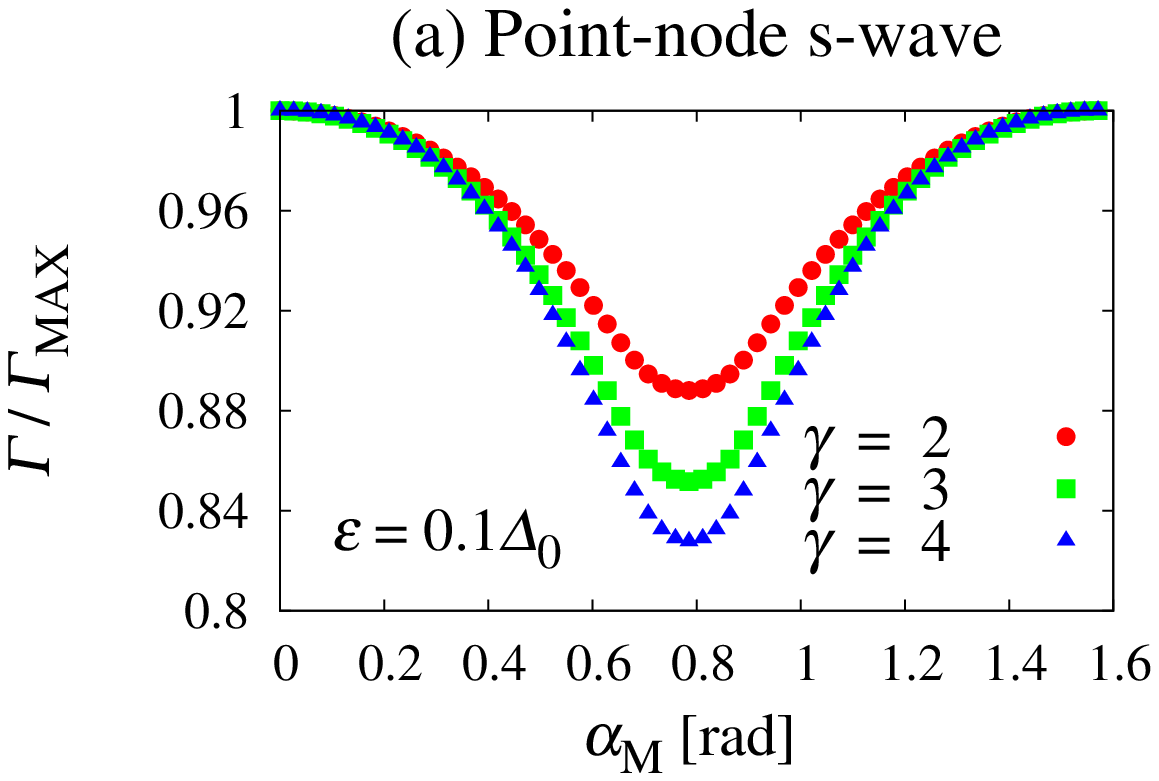}}\\
      \resizebox{80mm}{!}{\includegraphics{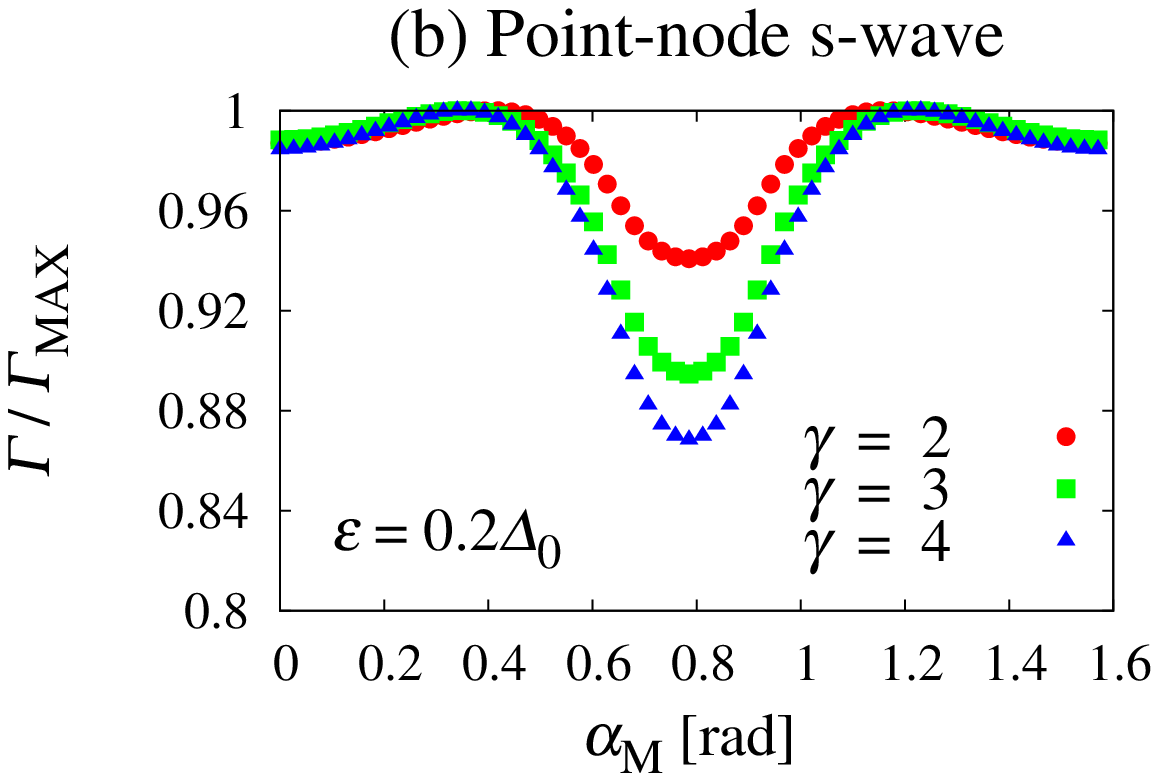}}
    \end{tabular}
 \end{center}
 \caption{Plot of the QP scattering rate $\varGamma$ vs.\ the applied field angle $\alpha_{\rm M}$
in the case of the point-node $s$-wave pair for (a) the QP energy $\varepsilon=0.1\Delta_0$ and (b) $\varepsilon=0.2\Delta_0$.
Each curve is plotted for different anisotropy $\gamma$.
The vertical axis is normalized by maximum value for each curve.}
 \label{Fig.3}
\end{figure}
\begin{figure}[t]
 \begin{center}
   \begin{tabular}{p{80mm}p{80mm}}
     \resizebox{80mm}{!}{\includegraphics{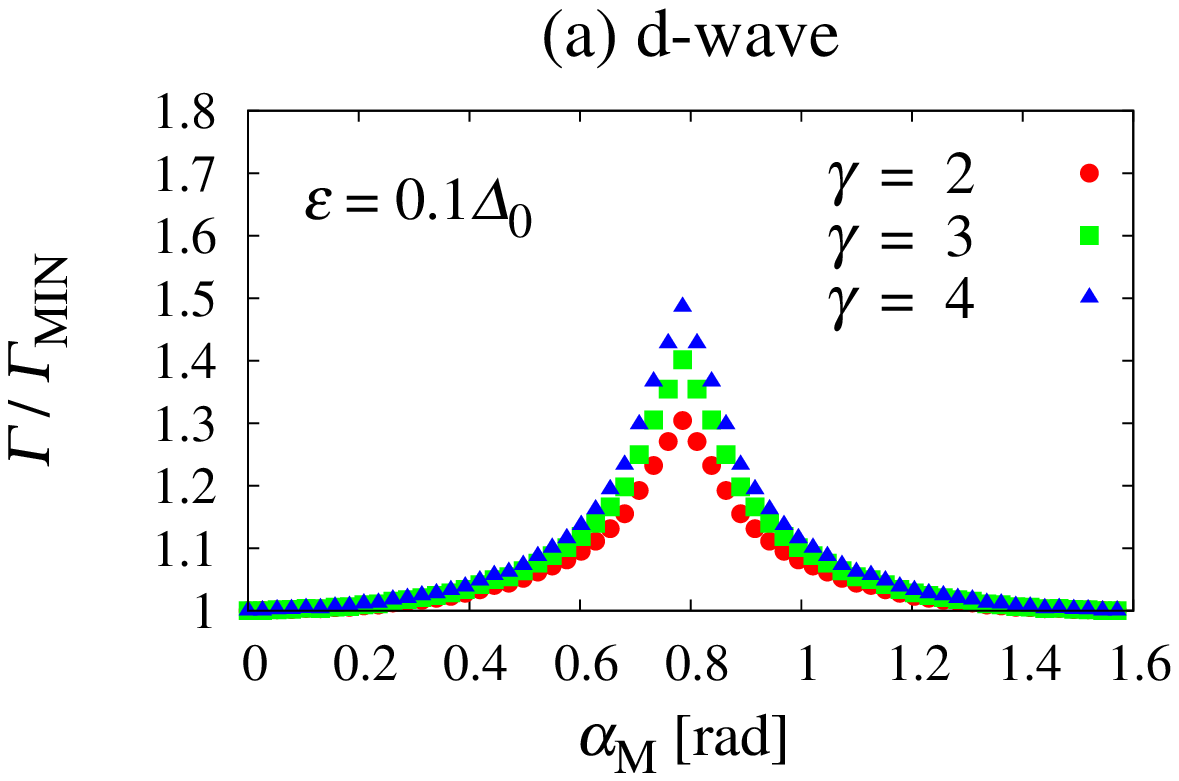}}\\
     \resizebox{80mm}{!}{\includegraphics{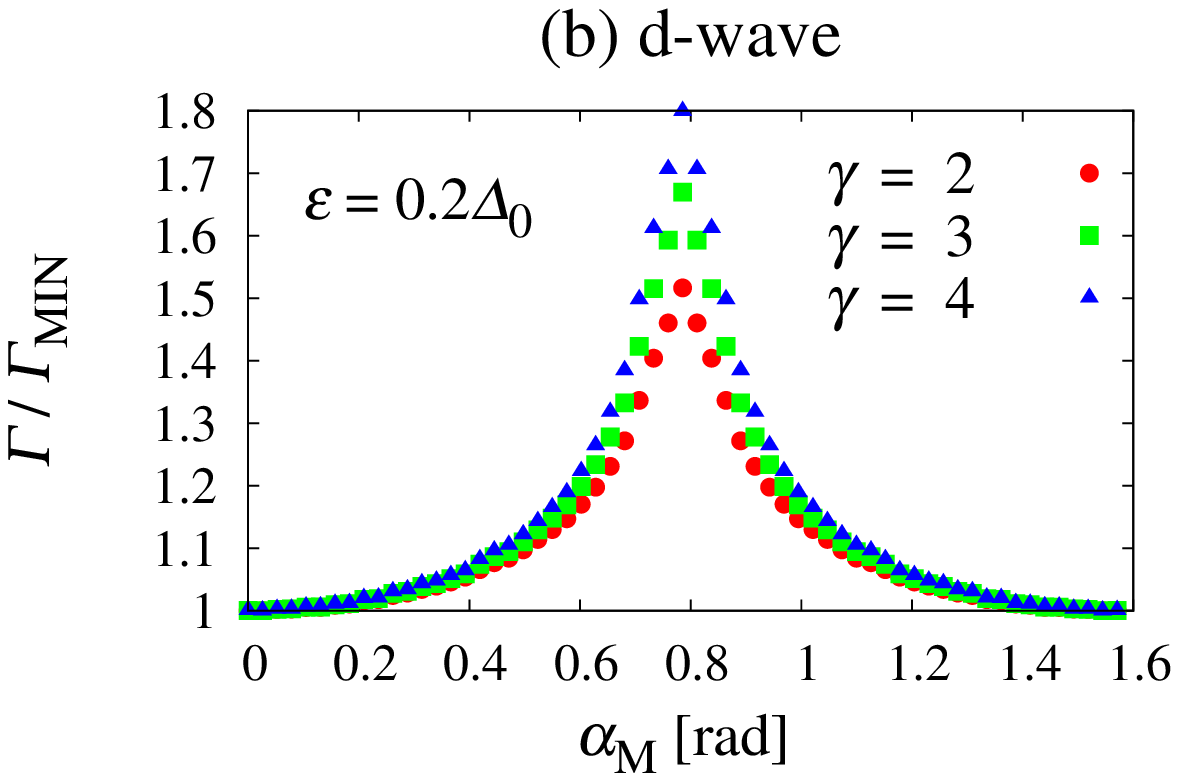}}
   \end{tabular}
 \end{center}
 \caption{Plot of the QP scattering rate $\varGamma$ vs.\ the applied field angle $\alpha_{\rm M}$
in the case of the $d$-wave pair for (a) the QP energy $\varepsilon=0.1\Delta_0$ and (b) $\varepsilon=0.2\Delta_0$.
Each curve is plotted for different anisotropy $\gamma$.
The vertical axis is normalized by minimum value for each curve.}
 \label{Fig.4}
\end{figure}

We consider the following three pair potential models on the spheroidal FS.
(i) Line-node $s$-wave:
$d({\bm k}_{\rm F})=| \cos 2\phi_k \sin^2\theta_k |$.
(ii) Point-node $s$-wave \cite{izawa}:
$d({\bm k}_{\rm F})=(1+\cos 4\phi_k \sin^4 \theta_k)/2$.
(iii) $d$-wave:
$d({\bm k}_{\rm F})=\cos 2\phi_k \sin^2\theta_k$.
These pairing functions have the gap nodes on the spheroidal FS
in the directions of $\phi_k=(1+2n)\pi/4$ with $n=0,1,2,3$.
The anti-node directions correspond to $\phi_k=n\pi/4$.
In Figs.~\ref{Fig.2}~-~\ref{Fig.4}, 
we show the $\alpha_{\rm M}$ dependence of $\varGamma$.
Each plot corresponds to the different $\gamma$.
In each pairing state, the left-side panel shows the numerical result for the QP energy $\varepsilon=0.1\Delta_0$
and the right-side one shows the result for $\varepsilon=0.2\Delta_0$.

As shown in Fig.~\ref{Fig.2}, in the case of the line-node $s$-wave pair,
$\varGamma$ exhibits its minimum when ${\bm H}$ is oriented parallel to the gap-node direction $(\alpha_{\rm M}=\pi/4)$.
With increasing $\gamma$, the minimum of $\varGamma$ gets smaller.
This behavior stands out in the higher energy.
In an isotropic spherical FS,
the oscillation amplitude of $\varGamma$ is of the order $6\%\ (7\%)$ of the maximum value for $\varepsilon=0.1\Delta_0\ (\varepsilon=0.2\Delta_0)$ \cite{higashi}.
On the other hand,
in the spheroidal FS with $\gamma=4$,
the amplitude is of the order $30\%\ (50\%)$.

While the similar behavior is seen also in the case of the point-node $s$-wave pair,
the difference appears in the energy dependence of $\varGamma(\alpha_{\rm M})$.
In the line-node $s$-wave case,
the minimum of $\varGamma(\alpha_{\rm M})$ gets smaller with increasing the QP energy $\varepsilon$.
However in the point-node $s$-wave case,
it gets larger with increasing $\varepsilon$.
The other difference is seen in the oscillation pattern.
In the case of the line-node $s$-wave pair,
the cusp like structure appears as shown in Fig.~\ref{Fig.2},
but in the case of the point-node $s$-wave,
the cusp like structure disappears
and the broad minimum of $\varGamma(\alpha_{\rm M})$ appears (see Fig.~\ref{Fig.3}).

The similar calculations are conducted also for the $d$-wave pair (Fig.~\ref{Fig.4}).
The maximum of $\varGamma$ appears when ${\bm H}$ is parallel to the node direction.
With increasing $\gamma$, the peak of $\varGamma$ becomes higher.
In an isotropic system,
the peak height is of the order $120\%\ (140\%)$ of the minimum value \cite{higashi}.
On the other hand,
in the spheroidal FS with $\gamma=4$,
the peak height is of the order $150\%\ (180\%)$.

In summary, the difference in $\varGamma(\alpha_{\rm M})$ between the $s$-wave and the $d$-wave pair is also seen for the spheroidal FS.
With increasing the two dimensionality of the Fermi surface, the difference becomes prominent.

\section{Discussion and Conclusion}
We clarify that
$\varGamma(\alpha_{\rm M})$ reflects 
a sign-change of the pair potential
more clearly in a
spheroidal FS
than in an isotropic spherical FS.
With increasing $\gamma$,
the difference of the behavior of $\varGamma(\alpha_{\rm M})$ between the $s$-wave and the $d$-wave pair
becomes larger.
Therefore, $\varGamma(\alpha_{\rm M})$ is phase-sensitive also for a spheroidal FS.
Information on the uniaxial anisotropy of FS is included in ${\bm v}_{\rm F \perp}(\bm{k}_{\rm F})$ in Eq.~(\ref{eq:1}).
The different behavior of $\varGamma(\alpha_{\rm M})$ between a spherical and a spheroidal FS
comes from the different magnitudes of ${\bm v}_{\rm F \perp}({\bm k}_{\rm F})$ and ${\bm v}_{\rm F \perp}({\bm k}^\prime_{\rm F})$.

In moderately clean regime, the flux-flow resistivity $\rho_{\rm f}(T)$ is proportional to $\varGamma(\varepsilon=k_{\rm B}T)$ \cite{kato}.
One can obtain the information of $\varGamma(\alpha_{\rm M})$ by measuring $\rho_{\rm f}$ with rotating $\bm{H}$ within the $a - b$ plane.
However, the strongly two dimensional system is not suitable for the measurment that we assume here,
since if $\bm{H}$ is applied within the $a - b$ plane, a Josephson vortex is formed \cite{yasuzuka},
which is beyond our theoretical framework.
In this work,
we consider an isotropic vortex core.
Elucidation of the effect of a vortex core distortion on $\varGamma$ is left for a future problem.
\section*{Acknowledgments}
We would like to thank N. Nakai, H. Suematsu, T. Okada, S. Yasuzuka, Y. Kato, K. Izawa, M. Kato, and A. Maeda for helpful discussions
and also thank N. Schopohl for sending the offprint of Ref. \cite{graser}.
\bibliographystyle{elsarticle-num}
\bibliography{<your-bib-database>}




\end{document}